# ESTIMATIONS OF LOCAL THERMAL IMPACT ON LIVING ORGANISMS IRRADIATED BY NON-THERMAL MICROWAVES

## Vladimir Shatalov[*]

*Department of Biophysics, Donetsk National University, Donetsk, Ukraine*

*Pennes'* differential equation for bioheat transfer and the heat transfer equation are solved for the temperature distribution in a living tissue with spherical inclusions, irradiated by microwave power. It is shown that relative temperature excess in a small inclusion in the tissue in some cases is inversely proportional to its radius and does not depend on the applied power. In pulsing RF fields the effect is amplified proportionally to the ratio of the pulse period to the pulse duration. The local temperature rise significantly outpaces the averaged one and therefore the Watt to Weight SAR limits may be insufficient to estimate the safety of RF radiation and the conventional division of the biological effects of electromagnetic fields on the thermal and non-thermal needs to be revised.

## INTRODUCTION

Traditionally, the effects of electromagnetic fields (EMF) on living organism are divided into ionizing (non-ionizing) and thermal (non-thermal). They are extensively investigated to date [e.g., NIEHS, 1999]. However, when considering effects of weak EMF, an important problem is to find and validate clear criteria of weakness of the impact [Challis, 2005; Shatalov, 2012]. For thermal effects of the EMF the presently accepted criteria are based upon application of physiologically acceptable limits to the increase of the *average* temperature of the samples.

Non-thermal effects defined in [ICEMS, 2010] are biological mechanisms that are not connected to the temperature increase, which is assumed to be less than 0.01degrees C (living organism), 0.001(cells) or 0.0005 (sub-cellular). By comparison, ANSI, WHO, IEEE & ICNIRP consider that exposures below 0.05 degrees C (0.4W/kg) are safe for workers, and exposures below 0.01 C (0.08 W/kg) are negligible for the public. Any biological effects below these levels of heating are considered by these organizations to have no biological significance and to be reversible. These criteria are used for estimation of maximum permissible power of domestic radio equipment (cell phones, Bluetooth, Wi-Fi, etc.). In particular, the effect of mobile phone radiation on human health is the subject of recent interest and study [Vecsei et al., 2013] due to the enormous increase in mobile phone usage throughout the world. Any biological effect of environmental RF EMFs is usually referred as non-thermal.

---
[*] Correspondence to: V.M. Shatalov, Biological Faculty, DonNU,
Schorsa st., 46, Donetsk, 83050, Ukraine
E-mail: v.shatalov@donnu.edu.ua

Another important application of these criteria lies with the development of non-lethal microwave weaponry (such as «Active Denial System», developed by the U.S. Department of Defense) to evaluate whether their effects are indeed reversible.

One well-understood effect of microwave radiation is dielectric heating, in which any polar solution (such as living tissue) is heated by rotations of the polar molecules induced by the electromagnetic field. In the case of a person using a cell phone, most of the heating effect occurs at the surface of the head, causing its temperature to increase by a fraction of a degree. The maximum power output from a mobile phone is regulated by the mobile phone standard and by the regulatory agencies in each country. In the USA, the Federal Communications Commission (FCC) has set a Specific Absorption Rate (SAR) limit of 1.6 W/kg, averaged over a volume of 1 gram of tissue, for the head. In Europe, the limit is 2 W/kg, averaged over a volume of 10 grams of tissue. SAR values are heavily dependent on the size of the averaging volume.

Non-thermal effects can also arise due to the low-frequency pulsing of the mobile phones carrier signal. Biological significance of these modulations is subject to a recent debate [Foster and, Repacholi, 2004]. Some researchers argue that so-called "non-thermal effects" can be reinterpreted as a normal cellular response to an increase in temperature. Others believe the stress proteins are unrelated to thermal effects, since they occur for both extremely low frequencies and radio frequencies (RF), which have very different energy levels [Blank and Goodman,. 2009]. Another study that was conducted using fluorodeoxyglucose injections and positron emission tomography concluded that exposure to radiofrequency signal waves within parts of the brain closest to the cell phone antenna results in increased levels of glucose metabolism, but the clinical significance of this finding is unknown [Volkow et al., 2011].

The aim of our paper is to give a mathematical model of the local thermal effects of the EMF by specifically taking into account inhomogeneity of the heating (resulting from non-uniformity of living tissue). The idea of the non-uniform heating, of course, is not new. However, that is hard to find specific estimates, therefore it was suggested by Challis [2005] as an open problem. Our main goal is to provide such analytical estimations and to give physical background of thermal effects related to inhomogeneity in space and time. We show that in some cases, the local temperature rise significantly outpaces the average absorbed radiation power and therefore the Watt to Weight SAR limits may be insufficient. The possibility of such a local heating (with the negligible average temperature increase) challenges the common division of EMF effects into thermal and non-thermal. While the effects, connected to the local heating, are thermal, the average heating may be negligible, making them classified conventionally as non-thermal.

Our consideration is not limited to any particular frequency band of the electromagnetic field. It is only important that the field is heating the conducting medium locally, be it via the eddy currents in the medium of non-uniform conductance, the polarization-induced vibrations of the molecules in the medium of non-uniform polarizability, or any other local mechanism. There are several different mechanisms, which lead to RF power absorption and have different penetration depths in different media.

It is obvious that the temperature distribution in any bounded sample heated by penetrated radiation has to be inhomogeneous. Firstly, it is due to the spatially inhomogeneous heating by the exponentially decaying RF power. It is well known that the temperature distribution has maximum under the sample surface if the heat transfer inside exceeds that one outside the sample (for example, due to pure transparency of the sample surface). The maximum value is limited by RF power decay depth that actually varies from fractions of a millimeter to several centimeters. Therefore, this maximum is flat and does not exceed much the surface temperature value. Unlike the case described, we will examine the temperature peaks at the micro inclusions having high electrical conductivity, which are located far enough away from the cooling surfaces or heat sinks.

## TEMPERATURE DISTRIBUTION IN THE CASE OF RANDOMLY DISTRIBUTED POINT SOURCES

It is well known that living tissue is substantially heterogeneous on micro and nano-scales with some of its components having increased electrical conductivity. For example: 1) electrical conductivity of the cytoplasm of living cells is somewhat higher than that of extracellular media due to the Na-K asymmetry [Lyashchenko and Lileev, 2010]; 2) electric double layer, forming on the surface of colloidal inclusions (such as air nano-bubbles in water), consists of a diffuse cloud of ions with superior conductivity compared to that of the solvent [Shatalov et al., 2012]; 3) axon membranes have extra high conductivity, etc. There are also numerous instances of nano-scale conductance inhomogeneity in inanimate nature exemplified by any colloidal solution (liquid or solid) of conductive nanoparticles in weakly conductive medium. Since the absorbed microwave and RF power is proportional to the electrical conductivity, conducting inclusions are heated by radiation more intensively than the surrounding media.

To describe this process of non-uniform heating in the medium with thermal regulation (present in all living tissue on all scales) let us use the classical *Pennes'* differential equation [Pennes, 1948]:

$$\frac{\partial \Theta}{\partial t} = k\nabla^2 \Theta - \frac{\Theta}{\tau} + \frac{W}{C_V}, \qquad (1)$$

where $\Theta$ is temperature excess [K] above the "normal" temperature, $t$ – time [s], $k$ – temperature conductivity [m² s⁻¹] (for example, in water $k$ =0.15mm²/s at normal conditions), $\nabla^2$ – Laplace operator, $\tau$ is characteristic temperature relaxation time [s], $W$ – absorbed power [J m⁻³ s⁻¹], $C_v$ – volumetric heat capacity [J/(m³ K)]. The second term on the right hand side describes thermal regulation, whatever the mechanism is. In the absence of heat sources (W=0) the medium relaxes towards zero temperature excess ($\Theta$), or, to its "normal" temperature. In steady state, when time derivative is zero, Eq. (2) reduces to the well-known *Helmholtz* equation (or screened *Poison* equation):

$$\nabla^2 \Theta(\mathbf{r}) - \frac{\Theta(\mathbf{r})}{R_\tau^2} = -q(\mathbf{r}), \tag{2}$$

where $R_\tau = (k\,\tau)^{1/2}$ is a characteristic size of the temperature relaxation region, $q = W/C_V k$ – renormalized absorbed heat [K/m²]. This equation is linear and for an arbitrary $q(\mathbf{r})$ its solution can formally be written with the help of Green's function as

$$\Theta(\mathbf{r}) = \iiint_V q(\mathbf{r}) \frac{e^{-|\mathbf{r}-\mathbf{r}'|/R_\tau}}{4\pi |\mathbf{r}-\mathbf{r}'|} d^3\mathbf{r}'. \tag{3}$$

Next we assume that heat is supplied in the form of randomly distributed point sources of the density $n$ and power $W$. The average temperature excess in the sphere of radius $R_i$ around each source can be obtained by direct integration as

$$\Theta_i = \frac{3W\tau}{4\pi C_V R_i^3}\left(1 - e^{-x} - x e^{-x}\right), \quad x = R_i / R_\tau. \tag{4}$$

The average equilibrium temperature excess, measured in the whole medium, is then

$$\Theta_0 = \frac{W n \tau}{C_V}, \tag{5}$$

Assuming that temperature conductivity of the media and inclusions are the same and identifying $R_i$ with inclusion size, the *relative* temperature excess in a small inclusions ($x \ll 1$) is inversely proportional to its radius and does not depend on the input power:

$$\frac{\Theta_i}{\Theta_0} = \frac{3}{8\pi\, k\, n\, \tau\, R_i}. \tag{6}$$

For small low-density inclusions in a living tissue with weak thermal regulation and low temperature conductivity, this ratio can be extremely high. Furthermore, if the heating is not stationary but consists of pulses of the width $\tau_w$ and the period $\tau_p > \tau_w$, it is easy to show that the local heating during the pulse can also be represented by (6) with an additional factor of $\tau_p / \tau_w$. That is, the pulsing EMF source leads to even higher local increase of the relative temperature.

# HEATING OF A SINGLE SPHERICAL INCLUSION IN A SPHERICAL SAMPLE

*Pennes'* equation (1) assumes that the total energy exchange between tissue and the flowing blood can be modeled as a non-directional heat outlet, whose magnitude is proportional to the volumetric blood flow and the difference between local tissue and major supply arterial temperatures. This approximation is valid for large enough samples, containing numerous blood vessels. To investigate the temperature distribution at a smaller scale in this section, we explore a different model, in which a heated center indirectly interacts with the cooling surface. Such a model might describe the temperature distribution in a close vicinity of a blood vessel that we refer as a thermostat. To simplify the consideration, the living tissue will be modeled as a sphere. Starting from such a spherical horse, we mean to explain the origin of temperature spikes in nano-inclusions contained in any tissue of any form.

## Stationary temperature distribution

Let us consider a spherical inclusion of radius $R_i$ in the center of the irradiated spherical sample with radius $R_0$, placed in a thermostat. The inclusion and the surrounding media have different electrical and the same temperature conductivities. The stationary equation of heat transfer in this case looks like:

$$\frac{\partial^2 \Theta(r)}{\partial r^2} + \frac{2}{r} \frac{\partial \Theta(r)}{\partial r} + q(r) = 0 \quad . \tag{7}$$

Here *r*-dependence of the renormalized absorbed heat has the form:

$$q(r) = \begin{cases} q_i, & 0 < r < R_i \\ y q_i, & R_i < r < R_0 \end{cases}, \tag{8}$$

where factor *y* points to the abovementioned difference of electrical properties inside and outside the inclusion.

Obviously, any solution of (7) has the form:

$$\Theta(r) = \frac{A}{r} - \frac{B}{6} r^2 + C \quad . \tag{9}$$

where the constants *A*, *B*, and *C* differ for the inclusion and outer space. Putting (9) into (7) we get $B_i = q_i$ and $B_0 = y q_i$. Then from the condition of zero flow at $r=0$ we get $A_i = 0$. Next, the flow continuity condition at the boundary of the inclusion gives the value of $A_0$, the value $C_0$ we get by putting $\Theta(R_0)=0$. Finally, the equality of temperatures at the boundary of the inclusion gives us the value $C_i$. Thus we get the solution of (7), that allows us to calculate the average temperature of the inclusion $\Theta_i$ and the outer space $\Theta_0$ and its ratio:

$$\frac{\Theta_i}{\Theta_0} = \frac{(1+x+x^2)}{(1-x)} \frac{[12x^2 - 10x^3 + 5y(1-3x^2+2x^3)]}{[5x^3+10x^4+2y(1+2x+3x^2-x^3-5x^4)]}, \quad (10)$$

where $x = R_i/R_0$ and $y = q_0/q_i$. For the case under interest $x \ll 1$ and $y=0$ we get:

$$\frac{\Theta_i}{\Theta_0} = \frac{12}{5} \frac{R_0}{R_i}. \quad (11)$$

So, the ratio of average temperatures of the inclusion to that one of the outer space is reversly proportional to its radii. This qualitatively coincidences with (6) outlined in previous section and differs just in coefficient.

In the case of a homogeneous medium, when properties of the inclusion and the surrounding environment are the same ($y=1$), then for $x \ll 1$ we get:

$$\frac{\Theta_i}{\Theta_0} = \frac{5}{2}. \quad (12)$$

Thus, the temperature of a micro-inclusion exceeds the average temperature 2.5 times just because the inclusion is far enough from the thermo-stabilized outer boundary of the sample.

Finally, we may conclude that there exist two reasons giving temperature increase in a little inclusion. First is the increased heat production, and second – the remote distance from the cooling boundary.

### Non-stationary temperature distribution

Another source of the temperature non-homogeneity is time inhomogeneous heating. In this subsection, we explore the non-stationary case for the spherical model from previous section.

At the starting point, a short RF pulse heats the spherical inclusion and surrounding media and we monitor the temperature distribution in space and time $\Theta(r,t)$. To get the distribution we solve numerically the non-stationary equation of heat transfer (1) for a simple case of two co-centered spherical media with different electrical and the same temperature conductivities. We are interested in the time range $t \ll \tau$ therefore the term that describes thermal regulation in (1) may be omitted. So, in the spherical coordinates centered in the cell center the equation (1) takes the form:

$$\frac{\partial \Theta(r,t)}{\partial t} = \frac{\partial^2 \Theta(r,t)}{\partial r^2} + \frac{2}{r} \frac{\partial \Theta(r,t)}{\partial r} + Q(r,t), \quad (13)$$

where $r$ and $t$ are given in units $R_0$ and $\tau = R_0^2/k$ correspondently. We state the absorbed power distribution $Q(r,t) = R_0^2 q(r,t)$ as

$$q(r,t) = \begin{cases} q_i \sin\left(\dfrac{\pi t}{\tau_w}\right), & (0 \le r \le R_i) \cap (0 \le t \le \tau_w) \\ q_0 \sin\left(\dfrac{\pi t}{\tau_w}\right), & (R_i < r) \cup (0 \le t \le \tau_w) \\ 0, & (\tau_w < t) \end{cases} \qquad (14)$$

where $x = R_i/R_0$ and $\tau_w$ is the RF pulse duration time, $q_i$ and $q_0$ – heated power inside and outside of the inclusion, its ratio is proportional to the ratio of correspondent electric conductivities. The initial and boundary conditions for the eq. (13) have the form:

$$\begin{cases} \Theta(r,0) = 0, & 0 < r < \infty; \\ \dfrac{\partial \Theta(r,t)}{\partial r} = 0, & r = 0; \\ \Theta(R_0,t) = 0. \end{cases} \qquad (15)$$

The results of our calculations of the relative temperature dynamics in a sample irradiated by a single RF pulse with $\tau_w = 0.2\tau$ for the size ratio $x=0.2$ and different heated power ratios $y = 0, 1, 2, 4$ and $\infty$ are shown in Figure 1.

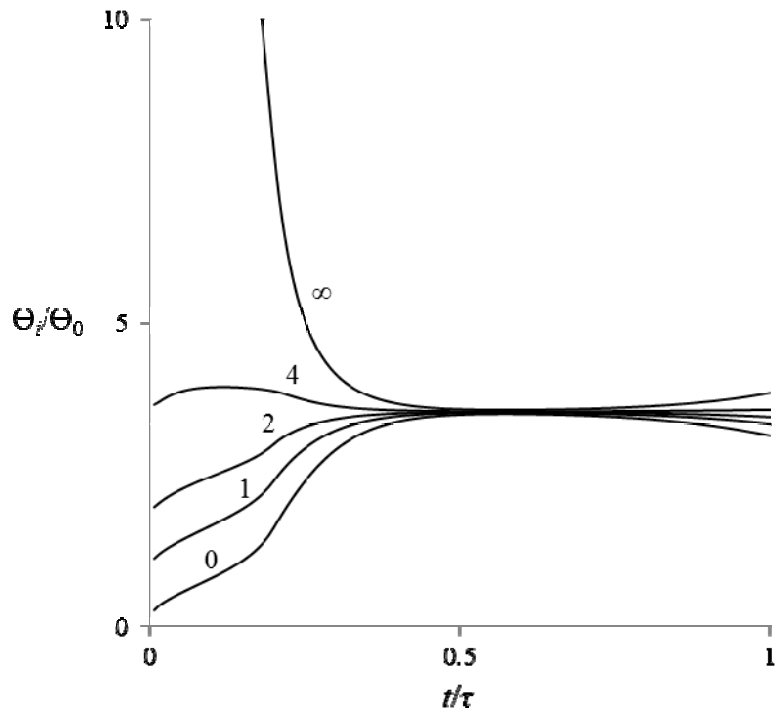

Figure 1. Relative average temperature of the inclusion with $R_i = 0.2R_0$ as a function of time (in units of $\tau$) in a sample irradiated by RF pulse (14) with $\tau_w = 0.2\tau$. Numbers over the curves are $q_i/q_0$ ratios of RF power absorption inside and outside the inclusion.

In the case of dielectric inclusion in conducting media ($y = 0$) the temperature of media exceeds the temperature of inclusion $\Theta_i/\Theta_0 < 1$ only at initial moments, but soon after this ratio rapidly increases to some asymptotic value close to 3. For larger $y$ the ratio $\Theta_i/\Theta_0$ exceeds 1 at all times. For the conducting inclusion in dielectric media ($y = \infty$) the relative temperature $\Theta_i/\Theta_0$ is very high in initial moments because the media is heating via heat transfer from the inclusion. After the heating pulse finished all the curves go to the asymptotic value.

The curves on Figure 1 were calculated for the case of a single RF pulse but it is obvious if the time scale expands to a sequence of pulses then every next pulse will start at increased values of relative temperature of the inclusion and that amplifies the effect of inhomogeneous heating of the sample.

## CONCLUSIONS

The results obtained in the paper have to draw attention to the obvious and at the same time unexpectedly large spatial inhomogeneity of heating of the irradiated samples, regardless of the intensity of irradiation. Until now, it seemed to be obvious that RF power is the main criteria, which refers any RF exposure to the thermal or non-thermal one. However, we have shown that relative temperature rise in micro inclusions is RF power independent and may be very high because it is reversely proportional to the size of the inclusions. This is the case when the radiation is absorbed mainly in the inclusions that situated sufficiently far from the heat outlets. Moreover, qualitatively this conclusion is an obvious consequence from Fourier's law and the heat transfer boundary conditions, and this is a long-solved problem in physics that leaves no questions.

To prove the results experimentally one has to measure the local temperature inside nano-objects that is not so easy but may be very informative. For example, by coating their samples with molecular layers with well-defined melting temperatures, Samsonov and Popov, [2013] measured the temperature in the immediate vicinity of individual cells during exposure to the EMF. They detected a rapid temperature jump in the experimental chamber and rapid establishment of the steady-state temperature. This allowed them to compare, in a quantitative manner, the cellular effects of the EMF with those of the temperature jump elicited by conventional heating.

There exist many experimental works with inanimate samples, the results of which, we believe, can be interpreted in terms of local heating (or overheating) of irregularities in the irradiated samples. For example, Bunkin et al., [2010] very carefully examined the effect of dissolved gases on some of the properties of distilled water. They showed that in presence of dissolved air IR laser irradiation heats the microbubbles of air to plasma temperatures that increases absorption of the radiation, while in the absence of dissolved gases, the water is essentially transparent due to low probability of the multiphoton absorption mechanism. In the work of

Doroshkevych et al. [2012] it was shown that weak pulsed magnetic field speed up self-organization processes in the nanopowder dispersed systems based on compacted $ZrO_2$. The list of examples may be easily continued. We believe that all these effects may be consequences of the inhomogeneous heating in micro- and nano-scales.

There exists a lot of contradictory evidence for non-thermal effects of electromagnetic radiation on living organisms. Some of them were pointed in Introduction [Challis, 2005; Shatalov, 2012; ICEMS, 2010]. According to our paper, many of the interpretations need to be revisable. So, we are confident enough to judge the dispute in the works Foster and Repacholi, [2004]; Blank and Goodman, [2009] in favor of the thermal origin of the observed effects. Another example, D'Andrea et al., [2003] pointed that effects of low-level RF exposure on the blood–brain barrier are controversial and the effects have been generally accepted for exposures that are thermal. Studies at these levels have observed effects on norepinephrine, dopamine, and serotonin. Now we can say that thermal mechanisms have to be favorable in explanation of any effect of MW on neurochemistry. Concerning discussions in Internet on safety of the non-lethal microwave weapon, it should be pointed to possible uncontrolled temperature rise in under skin inclusions.

So, the local temperature can rise significantly outpaces the averaged one and therefore the Watt to Weight SAR limits may be insufficient to account the safety of RF radiation and the conventional division of the biological effects of electromagnetic fields on the thermal and non-thermal needs to be revised.

## ACKNOWLEDGMENTS

The author thanks Prof. Andrew Lyaschenko for discussions that served as a stimulus to the writing this article. He is also very grateful to Dr. Konstantin Metlov for reading and discussing the manuscript and to Prof. Nicholas Bunkin for helpful comments.

## REFERENCES

Blank M, Goodman R. 2009. Electromagnetic fields stress living cells. Pathophysiology 16 (2–3): 71–78.
Challis LJ. 2005. Mechanisms for interaction between RF fields and biological tissue. Bioelectromagnetics; Supplement 7:S98-106.
D'Andrea JA, Chou CK, Johnston SA, Adair ER. 2003. Microwave effects on the nervous system. Bioelectromagnetics Supplement. 6:S107-S147.
Doroshkevych OS, Shylo AV, Saprukina OV, Danilenko IA, Konstantinova TE, Ahkozov LA. 2012. Impedance spectroscopy of concentrated zirconia nanopowder dispersed systems experimental technique. World Journal of Condensed Matter Physics. 2(1):1-9.
Foster KR, Repacholi MH. 2004. Biological effects of radiofrequency fields: does modulation matter? Radiation Research 162 (2): 219–244.

Glaser R. 2005. Are thermoreceptors responsible for "non-thermal" effects of RF fields? Ed. Wissenschaft (Bonn, Germany: Forschungsgemeinschaft Funk) (21). OCLC 179908725.

ICEMS. 2010. Non thermal effects and mechanisms of interaction between electromagnetic fields and living matter. ICEMS, eds. Guiliani L. & Soffritti M.: Ramazzini Institute, European J of Oncology, Library Vol. 5.

Lyashchenko AK, Lileev AS. 2010. Dielectric relaxation of water in hydration shells of ions. J. Chem. Eng data. 55:2008-2016.

Bunkin NF, Ninham BW, Babenko VA, Suyazov NV, Sychev AA. 2010. Role of dissolved gas in optical breakdown of water: differences between effects due to helium and other gases. J. Phys. Chem. B 114:7743–7752.

NIEHS. 1999. Report on health effects from exposure to power-line frequency electric and magnetic fields. National Institute of Environmental Health Sciences of the U.S. National Institutes of Health. NIH Publication No. 99-4493.

Pennes HH. 1948. Analysis of tissue and arterial blood temperature in the resting human forearm. J. Appl. Physiol. 1:93–122.

Samsonov A, Popov SV. 2013. The effect of a 94 GHz electromagnetic field on neuronal microtubules. Bioelectromagnetics, 34:133–144.

Shatalov VM. 2012. Mechanism of the Biological Impact of Weak Electromagnetic Fields and the In Vitro Effects of Blood Degassing. Biophysics 57:808-813.

Shatalov VM, Filippov AE, Noga IV. 2012. Bubbles induced fluctuations of some properties of aqueous solutions. Biophysics. 57(4):421–427.

Vecsei Z, Csathó A, Thuróczy G, Hernádi I. 2013. Effect of a single 30 min UMTS mobile phone-like exposure on the thermal pain threshold of young healthy volunteers. Bioelectromagnetics. 34(7):530-541.

Volkow ND, Tomasi D, Wang G-J, Vaska P, Fowler JS, Telang F, Alexoff D, Logan J, et al. 2011. Effects of cell phone radiofrequency signal exposure on brain glucose metabolism. JAMA 305 (8): 808–821.